\documentclass[journal,12pt,draftclsnofoot,onecolumn]{IEEEtran}
\usepackage{mathrsfs}
\usepackage{cite}
\usepackage{amsmath}
\usepackage{amssymb}
\usepackage{stfloats}
\usepackage{graphicx}
\usepackage{setspace}
\usepackage{xcolor}

\ifCLASSOPTIONcompsoc
\usepackage[tight,normalsize,sf,SF]{subfigure}
\else
\usepackage[tight,footnotesize]{subfigure}
\fi

\begin{document}
\title{Quantization and Feedback of Spatial Covariance Matrix for Massive MIMO Systems with Cascaded Precoding}

\author{Yinsheng~Liu, Geoffrey Ye Li, and Wei Han.
\thanks{Yinsheng Liu is with State Key Laboratory of Rail Traffic Control and Safety, Beijing Jiaotong University, Beijing 100044, China, e-mail: ys.liu@bjtu.edu.cn.}
\thanks{Geoffrey Ye Li is with ITP lab, School of ECE, Georgia Institute of Technology, Atlanta 30313, Georgia, USA, e-mail: liye@ece.gatech.edu.}
\thanks{Wei Han is with Huawei Technologies, Co. Ltd., e-mail: wayne.hanwei@huawei.com.}
}

\maketitle
\doublespacing

\begin{abstract}
In this paper, we investigate the quantization and the feedback of downlink spatial covariance matrix for massive multiple-input multiple-output (MIMO) systems with cascaded precoding. Massive MIMO has gained a lot of attention recently because of its ability to significantly improve the network performance. To reduce the overhead of downlink channel estimation and uplink feedback in frequency-division duplex  massive MIMO systems, cascaded precoding has been proposed, where the outer precoder is implemented using traditional limited feedback while the inner precoder is determined by the spatial covariance matrix of the channels. In massive MIMO systems, it is difficult to quantize the spatial covariance matrix because of its large size caused by the huge number of antennas. In this paper, we propose a spatial spectrum based approach for the quantization and the feedback of the spatial covariance matrix. The proposed inner precoder can be viewed as modulated discrete prolate spheroidal sequences and thus achieve much smaller spatial leakage than the traditional discrete Fourier transform submatrix based precoding. Practical issues for the application of the proposed approach are also addressed in this paper.
\end{abstract}


\newpage
\section{Introduction}
Massive multiple-input multiple-output (MIMO) formed by installing a huge number of antennas at the base station (BS) can significantly increase the spectrum- and energy-efficiencies of wireless networks \cite{HQNgo,EGLarsson}. As a promising technique for the next generation cellular systems, it has gained a lot of attention recently \cite{LLu,FRusek}.\par
For the best performance gain of downlink transmission, accurate downlink channel state information (CSI) is required at the BS for precoding \cite{EGLarsson}. In a time-division duplex system, downlink CSI at the BS can be obtained by exploiting the channel reciprocity \cite{EGLarsson,FRusek,HQNgo2}. However, due to the huge number of antennas in massive MIMO systems, the acquisition of downlink CSI at the BS is not that easy for a frequency-division duplex (FDD) system even if most existing cellular networks are operated in FDD mode.\par
For traditional MIMO in FDD systems, downlink CSI can be estimated at the user equipment (UE) and fed back to the BS \cite{YLiu_survey,DJLove}. In massive MIMO systems, however, traditional channel estimation and limited feedback techniques can be hardly used because of the large overhead caused by the huge number of antennas. To address this issue, a closed-loop training technique has been proposed in \cite{JChoi2,AJDuly} for downlink CSI acquisition. It can achieve significant performance improvement within only a few iterations by exploiting the spatial correlation of the channels corresponding to different antennas. The channel spatial correlation has also been used in \cite{JChoi3,JChoi4,MSSim} to develop a trellis coded quantization in order to reduce the feedback overhead. In \cite{JNam,AAdhikary}, a cascaded precoding is developed where the precoder is divided into an outer precoder and an inner precoder. The outer precoder is based on a low-dimension effective channel and can be implemented using traditional channel estimation and limited feedback. The inner precoder is determined by the eigen-matrix of the channel spatial covariance matrix. When the antenna number is very large, the eigen-matrix is usually approximated by a discrete Fourier transform (DFT) submatrix with each column corresponding to a beam. Therefore, the DFT submatrix can be used as the inner precoder to avoid the need of the covariance matrix \cite{AAdhikary,DYing,JBrady}. Apparently, the best performance of the cascaded precoding can be achieved by choosing the strongest beams to form the inner precoder, which will cause a large feedback overhead. To reduce the overhead, the feedback is conducted with respect to a group of sub-sectors in \cite{DYing}, and each sub-sector can be viewed as a wide beam that covers multiple narrow beams. The wide-beam based approach is also widely used to construct a multi-resolution codebook, which usually has multiple codebook levels with each level corresponding to a different angular resolution \cite{JWang,AAlkhateeb}.\par

Although the eigen-matrix of the downlink spatial covariance matrix can be approximated by the DFT submatrix when the antenna number is very large, spatial leakage exists and thus the DFT submatrix based inner precoder fails to capture the channel power efficiently when the antenna number in practical systems is always finite. In this case, the inner precoder based on the quantized covariance matrix is expected to achieve better performance. For covariance matrix quantization, a Lloyd-type algorithm has been proposed in \cite{SGhosh} where the codebook is generated by training with a large number of sample covariance matrices. However, the approach there can be hardly used in massive MIMO systems due to the huge size of the covariance matrix. In this paper, we develop a spatial spectrum based approach for the quantization and the feedback of the spatial covariance matrix. The quantization is conducted with respect to the spatial spectrum, which is defined as the Fourier transform of the spatial correlation, to avoid the complicated matrix calculation. A codebook can be therefore constructed using the quantized covariance matrices with each one generated from a corresponding codeword spectrum. Our analysis shows that the proposed inner precoder can be viewed as modulated discrete prolate steroidal sequences (DPS) \cite{DSlepian,TZemen}, and thus achieve the smallest spatial leakage among all kinds of inner precoders. As a consequence, the BS can not only focus the transmit power on the direction of interest but also yield much smaller interference in undesired directions.\par

Practical issues regarding the the application of the proposed approach will be also addressed in this paper, including the multiuser scenario, signal-to-noise ratio (SNR) estimation, a multi-codeword feedback scheme, and the uniform-planar-array (UPA) antenna case. First, our analysis shows that the proposed approach can be also used when there are multiple users in a practical system even though it is designed with respect to the single-user case. Second, the UE needs to estimate the SNRs of all codewords in order to pick the best one from the codebook. For SNR estimation, multiple training symbols are required to improve the codeword resolution \cite{JWang,AAlkhateeb,SHur}, or identify multiple main beams \cite{SHan}. In this paper, we will develop a training scheme that can estimate the SNRs for all codewords simultaneously with only one training symbol. The multi-codeword feedback scheme is also proposed to handle the situation where the angle-of-departures (AoDs) are distributed within multiple clusters. Finally, the UPA case is discussed since it is more widely used in practical systems.

The rest of this paper is organized as follows. The system model is presented in Section II. The codebook design is shown in Section III and an insightful analysis is presented in Section IV. Practical issues are discussed in Section V. Simulation results are presented in Section VI, and conclusions are finally drawn in Section VII.

\section{System Model}
\begin{figure}
  \centering
  \includegraphics[width=5in]{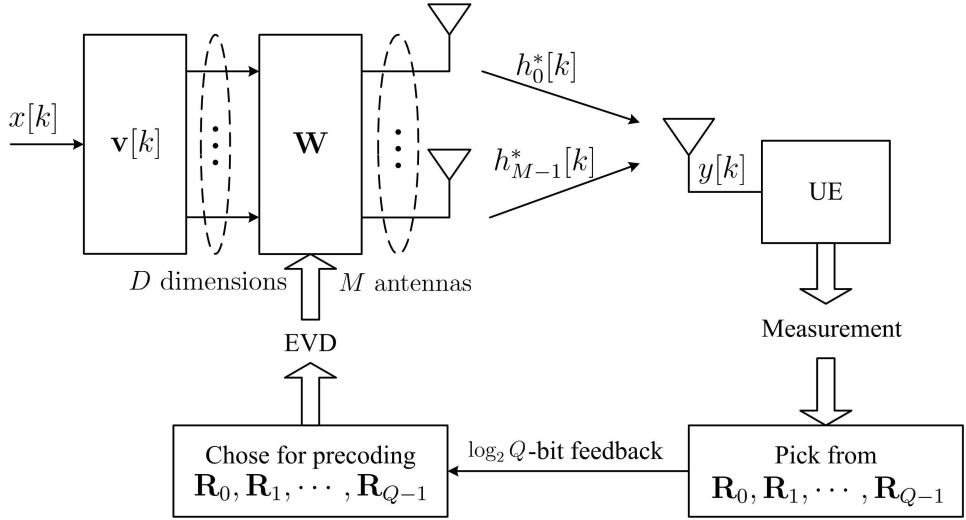}\\
  \caption{Cascaded precoding in downlink massive MIMO system with a codebook composed of quantized covariance matrices. UE determines the best quantized covariance matrix and feed back its index to the BS. For practical realization, $\mathbf{U}_q$'s can be pre-calculated and stored in the BS. The complexity can be therefore greatly reduced since no need for real-time EVD operation.}\label{fig:system}
\end{figure}
As in Fig.~\ref{fig:system}, we consider a cascaded precoding for downlink transmission in an orthogonal frequency division multiplexing (OFDM) based massive MIMO system with $M$ antennas at the BS and a single antenna at the UE of interest \cite{AAdhikary}. The codebook is designed with respect to the single-user case and will be generalized to the multiuser case in Section V. Denote $\mathbf{v}[k]=\{v_d[k]\}_{d=0}^{D-1}\in\mathcal{C}^{D\times 1}$ to be the outer precoder at the $k$-th subcarrier, where $D$ is the size of the outer precoder and is much smaller than the antenna number. Denote $\mathbf{W}\in\mathcal{C}^{M\times D}=(\mathbf{w}_0,\cdots,\mathbf{w}_{D-1})$ with $\mathbf{w}_d=\{w_d[m]\}_{m=0}^{M-1}$ to be the inner precoder determined by the spatial covariance matrix. Note that different subcarriers can share the same inner precoder because the spatial covariance matrix is independent of the channel frequency selectivity. UE will determine the index of the best quantized covariance matrix through specially designed training and then feed its index back to the BS.\par
From Fig.~\ref{fig:system}, the received signal at the $k$-th subcarrier can be expressed as
\begin{align}\label{equ:sys_model}
y[k]=\mathbf{h}^{\mathrm{H}}[k]\mathbf{W}\mathbf{v}[k]x[k]+z[k],
\end{align}
where $x[k]$ is the transmit symbol at the $k$-th subcarrier with zero mean and unit variance, $z[k]$ is the corresponding additive Gaussian noise with zero mean and noise power, $N_0$, and $\mathbf{h}[k]=\{h_m[k]\}_{m=0}^{M-1}\in\mathcal{C}^{M\times 1}$ denotes the spatial channel vector with $h_m[k]$ indicating the channel frequency response at the $k$-th subcarrier on the $m$-th antenna.\par
In massive MIMO systems, the antennas at the BS are usually tightly placed, leading to high spatial correlation of the channels corresponding to different antennas \cite{OEAyach}. To capture
the nature of the spatial correlation, a physical channel model in \cite{AMSayeed} can be used to describe each propagation path between the antenna at the BS and the UE individually. For a typical uniform-linear-array (ULA) antenna,
\begin{align}
h_m[k]=\sum_{l=0}^{L-1}\alpha_{l}e^{-j\frac{2\pi k \tau_l}{T}}e^{j2\pi m v_l},
\end{align}
where $\alpha_{l}$ is the attenuation of the $l$-th path with zero mean and $\mathrm{E}(\alpha_{l}\alpha_{p}^*)=\sigma_l^2\delta[l-p]$ with $\delta[\cdot]$ indicating the Kronecker delta function and $\sum_{l=0}^{L-1}\sigma_l^2=1$ for normalization, $\tau_l$ denotes the delay corresponding to the $l$-th path, $T$ denotes the OFDM symbol duration, and $v_l=\frac{d}{\lambda}\sin\theta_l$ denotes the wave number of the $l$-th path with corresponding AoD $\theta_l$ and antenna spacing $d/\lambda$ normalized by the wave length. Accordingly, the spatial correlation function is expressed as
\begin{align}\label{equ:sp_corr}
r[m]\triangleq\mathrm{E}(h_{n+m}[k]h_n^*[k])=\sum_{l=0}^{L-1}\sigma_l^2e^{j2\pi m v_l},
\end{align}
or in a matrix for as $\mathbf{R}\triangleq\mathrm{E}(\mathbf{h}[k]\mathbf{h}^{\mathrm{H}}[k])=\{r[m-n]\}_{m,n=0}^{M-1}$. Denote the eigen-value decomposition (EVD) of the covariance matrix, $\mathbf{R}$, as
\begin{align}\label{equ:corr_matrix}
\mathbf{R}=\mathbf{U}\mathbf{\Lambda}\mathbf{U}^{\mathrm{H}},
\end{align}
where $\mathbf{U}\in\mathcal{C}^{M\times M}$ denotes the eigen-matrix and $\mathbf{\Lambda}=\mathrm{diag}\{\lambda[m]\}_{m=0}^{M-1}$ is a diagonal matrix with $\lambda[m]$ indicating the $m$-th largest eigenvalue.\par

The outer precoder can be implemented using traditional limited feedback since it has a low dimension. For a given $\mathbf{R}$, the inner precoder is given by $\mathbf{W}=\mathbf{U}_D$, where $\mathbf{U}_D$ consists of the first $D$ columns of the eigen-matrix, $\mathbf{U}$, corresponding to the $D$ largest eigenvalues \cite{AAdhikary}. It is a challenging problem for the inner precoder design due to its large size caused by the huge number of antennas.\par

Since the inner precoder is determined by the spatial covariance matrix, a good codebook should be designed with respect to the quantization of the covariance matrix. Denote $\mathcal{R}=\{\mathbf{R}_0,\mathbf{R}_1,\cdots,\mathbf{R}_{Q-1}\}$ to be the codebook consisting of $Q$ quantized covariance matrices. With the specially designed training in Section V, the UE can determine the best codeword from the codebook $\mathcal{R}$. Each quantized covariance matrix, $\mathbf{R}_q=\{r_q[m-n]\}_{m,n=0}^{M-1}$, corresponds to a spatial correlation function, $r_q[m]$. If most power can be captured by the $D$ largest eigenvalues for each quantized covariance matrix, then the EVD of the quantized covariance matrix can be approximated by
\begin{align}\label{equ:evd_cw}
\mathbf{R}_q\approx\mathbf{U}_q\mathbf{\Lambda}_q\mathbf{U}_q^{\mathrm{H}},
\end{align}
where $\mathbf{U}_q=(\mathbf{u}_{q,0},\cdots,\mathbf{u}_{q,D-1})\in\mathcal{C}^{M\times D}$ with $\mathbf{u}_{q,d}=\{u_{q,d}[m]\}_{m=0}^{M-1}$ denotes the eigen-matrix with the columns corresponding to the $D$ largest eigenvalues\footnote{For practical realization, $\mathbf{U}_q$ can take the first $D$ columns of the eigen-matrix obtained from the EVD of $\mathbf{R}_q$.}, and $\mathbf{\Lambda}_q=\mathrm{diag}\{\lambda_q[m]\}_{m=0}^{D - 1}$ is a diagonal matrix with $\lambda_q[m]$ indicating the $m$-th largest eigenvalue of the $q$-th quantized covariance matrix. Therefore, designing the codebook is equivalent to finding $Q$ quantized covariance matrices.\par

\section{Codebook Design}
In this section, we will first introduce the design criterion based on the maximization of the average SNR, and then present the codebook design using spectrum quantization.
\subsection{Maximization of Average SNR}
To focus on the codebook design for the inner precoder, we assume an ideal outer precoder in this section, that is
\begin{align}
\mathbf{v}[k]=\frac{\mathbf{W}^{\mathrm{H}}\mathbf{h}[k]}{\|\mathbf{W}^{\mathrm{H}}\mathbf{h}[k]\|_2}.
\end{align}
In this case, the instantaneous SNR at the $k$-th subcarrier is given by
\begin{align}
\gamma[k]\triangleq\|\mathbf{W}^{\mathrm{H}}\mathbf{h}[k]\|_2^2,
\end{align}
where the noise power has been normalized for simple notation. The inner precoder is determined by the covariance matrix, which is long-term statistics of the channel. It thus should be designed with respect to the average SNR,
\begin{align}\label{equ:ave_snr}
\gamma\triangleq\mathrm{E}(\gamma[k])=\mathrm{Tr}(\mathbf{W}^{\mathrm{H}}\mathbf{R}\mathbf{W}),
\end{align}
rather than the instantaneous SNR, where $\mathrm{Tr}(\cdot)$ denotes the trace of a matrix.\par
If the $q$-th codeword is used, then $\mathbf{W}=\mathbf{U}_q$ and the average SNR for the $q$-th codeword is given by $\gamma_q= \mathrm{Tr}(\mathbf{U}_q^{\mathrm{H}}\mathbf{R}\mathbf{U}_q)$. To achieve the best performance, the UE should choose a codeword matrix to maximize the average SNR,
\begin{align}\label{equ:choose}
q^*=\operatorname*{arg~max}_{\{0,1,\cdots,Q-1\}} \gamma_q
\end{align}
Then, the index of the optimal codeword matrix, $q^*$, is fed back to the BS, which needs only $\log_2 Q$-bit feedback overhead.
\subsection{Spectrum Quantization}
As in (\ref{equ:choose}), the codeword is chosen to maximize the average SNR from a predetermined codebook. However, the codebook design based on (\ref{equ:choose}) is difficult, we therefore consider a bound expression, rather than the original average SNR, for the codebook design.\par

From Appendix A, we have
\begin{align}\label{equ:upper_bd}
\lambda_{q,\mathrm{max}}^{-1}\mathrm{Tr}(\mathbf{R}_q\mathbf{R})\leq\mathrm{Tr}(\mathbf{U}_q^{\mathrm{H}}\mathbf{R}\mathbf{U}_q)\leq\lambda_{q,\mathrm{min}}^{-1}\mathrm{Tr}(\mathbf{R}_q\mathbf{R}),
\end{align}
where $\lambda_{q,\mathrm{min}}$ and $\lambda_{q,\mathrm{max}}$ denote the minimum and the maximum eigenvalues of the $q$-th codeword matrix, $\mathbf{R}_q$, and the equality holds when all eigenvalues of $\mathbf{R}_q$ are equal. Using the bound expression, $\mathrm{Tr}(\mathbf{R}_q\mathbf{R})$ in (\ref{equ:upper_bd}), the criterion in (\ref{equ:choose}) can be rewritten as
\begin{align}\label{equ:correlation}
q^*=\operatorname*{arg~max}_{\{0,1,\cdots,Q-1\}} \mathrm{Tr}(\mathbf{R}_q\mathbf{R}),
\end{align}
Direct calculation yields that
\begin{align}
\|\mathbf{R}_q-\mathbf{R}\|^2_{\mathrm{F}}=\mathrm{Tr}(\mathbf{R}_q^{\mathrm{H}}\mathbf{R}_q+\mathbf{R}^{\mathrm{H}}\mathbf{R}-\mathbf{R}_q^{\mathrm{H}}\mathbf{R}-\mathbf{R}^{\mathrm{H}}\mathbf{R}_q).
\end{align}
If we assume that $\mathrm{Tr}(\mathbf{R}_q^{\mathrm{H}}\mathbf{R}_q)$'s are the same for different codewords in the designed codebook, then (\ref{equ:correlation}) is equivalent to
\begin{align}\label{equ:F_norm}
q^*=\operatorname*{arg~min}_{\{0,1,\cdots,Q-1\}} \|\mathbf{R}_q-\mathbf{R}\|^2_{\mathrm{F}},
\end{align}
where we have used the identities $\mathbf{R}=\mathbf{R}^{\mathrm{H}}$ and $\mathbf{R}_q=\mathbf{R}_q^{\mathrm{H}}$. From (\ref{equ:F_norm}), the best codeword should be closest to the channel covariance matrix in terms of the Frobenius-norm based distance.\par
On the other hand, the spatial spectrum can be defined as the Fourier transform of the spatial correlation in (\ref{equ:sp_corr}) \cite{SHaykin}, that is
\begin{align}
R(v)&\triangleq \sum_{m=-\infty}^{+\infty}r[m]e^{-j2\pi m v}=\sum_{l=0}^{L-1}\sigma_l^2\delta(v-v_l),
\end{align}
where $\delta(\cdot)$ denotes the Dirac delta function\footnote{Dirac delta function, which is used for the continuous signal, is different from the Kronecker delta function, which is used for the discrete case.}. Note that the spatial spectrum is always positive for a stationary random process. Direct calculation yields that
\begin{align}\label{equ:int}
\|\mathbf{R}_q-\mathbf{R}\|_{\mathrm{F}}^2&=\sum_{m=-\infty}^{+\infty}|(r_q[m]-r[m])a[m]|^2\nonumber\\
&=\int_{-\frac{1}{2}}^{+\frac{1}{2}}|R_q(v) * A(v) - R(v) * A(v)|^2dv\nonumber\\
&=\int_{-\frac{1}{2}}^{+\frac{1}{2}}[R_q(v) * A(v)]^2-2[R_q(v) * A(v)][R(v) * A(v)]+ [R(v) * A(v)]^2 dv,
\end{align}
where $*$ denotes the circular convolution, $R_q(v)$ is the Fourier transform of $r_q[m]$, and $A(v)$ is the Fourier transform of $a[m]$ that is defined as
\begin{align}
a[m]=
\begin{cases}
\sqrt{M-|m|}, & |m|\leq M-1\\
0, & \mathrm{otherwise}
\end{cases}.
\end{align}
If we assume that the first term in the third equation of (\ref{equ:int}) is the same for all codewords in the designed codebook, then the criterion in (\ref{equ:F_norm}) can be rewritten as
\begin{align}
q^*&=\operatorname*{arg~max}_{\{0,1,\cdots,Q-1\}}\int_{-\frac{1}{2}}^{+\frac{1}{2}}[R_q(v) * A(v)][R(v) * A(v)]dv\nonumber\\
&=\operatorname*{arg~max}_{\{0,1,\cdots,Q-1\}}\int_{-\frac{1}{2}}^{+\frac{1}{2}}\int_{-\frac{1}{2}}^{+\frac{1}{2}}[R_q(v) * A(v)]A(v-\tau)R(\tau)dvd\tau\nonumber\\
&=\operatorname*{arg~max}_{\{0,1,\cdots,Q-1\}}\int_{-\frac{1}{2}}^{+\frac{1}{2}}[R_q(v) * G(v)]R(v)dv,
\end{align}
where we have used the identity $A(v)=A(-v)$ since $a[m]=a[-m]$, and
\begin{align}
G(v)=A(v) * A(v)=\frac{\sin^2(\pi M v)}{\sin^2 (\pi v)}.
\end{align}
If the antenna number is very large as in massive MIMO systems, then $G(v)\approx\delta(v)$. As a result, the best codeword should be determined by
\begin{align}
q^*=\operatorname*{arg~max}_{\{0,1,\cdots,Q-1\}}\int_{-\frac{1}{2}}^{+\frac{1}{2}}R_q(v)R(v)dv,
\end{align}
or equivalently,
\begin{align}\label{equ:chordal}
q^*=\operatorname*{arg~min}_{\{0,1,\cdots,Q-1\}}\sqrt{1-\left|\int_{-\frac{1}{2}}^{+\frac{1}{2}}R_q(v)R(v)dv\right|^2},
\end{align}
where the right side of the equation can be viewed as minimizing the chordal distance on a functional space that has infinite dimensions.\par

In \cite{KKMukkavilli}, a geometric model has been developed to solve the optimal codebook design for beamforming in regular MIMO systems, where the distance metric has a similar form as (\ref{equ:chordal}) except that the latter is defined on the finite dimension space. We can similarly adopt such a model for the codebook design in this paper. As a result, the design of good codebook turns to the following optimization problem,
\begin{align}\label{equ:max_min}
\operatorname*{max}_{\mathcal{R}}\operatorname*{min}_{0\leq p,q\leq Q-1}\sqrt{1-\left|\int_{-\frac{1}{2}}^{+\frac{1}{2}}R_p(v)R_q(v)dv\right|^2},
\end{align}
that is, the codebook should be designed to maximize the minimum chordal distance on a functional space between different codewords.\par

By viewing each codewrod as a point on a Grassmannian manifold, Grassmannian line packing has been used to design a good codebook, assuming that the codebook size is not smaller than the antenna number \cite{DJLove_Unitary}. However, such an approach cannot be used on a functional space because the dimension of the functional space is infinite and is therefore always larger than the codebook size.\footnote{Although the Grassmannian line packing cannot be used in this paper, the criterion for designing a good codebook in (\ref{equ:max_min}) is still valid because the geometric model in \cite{KKMukkavilli} can work when the antenna number is larger than the codebook size.}\par

\begin{figure}
  \centering
  \includegraphics[width=4in]{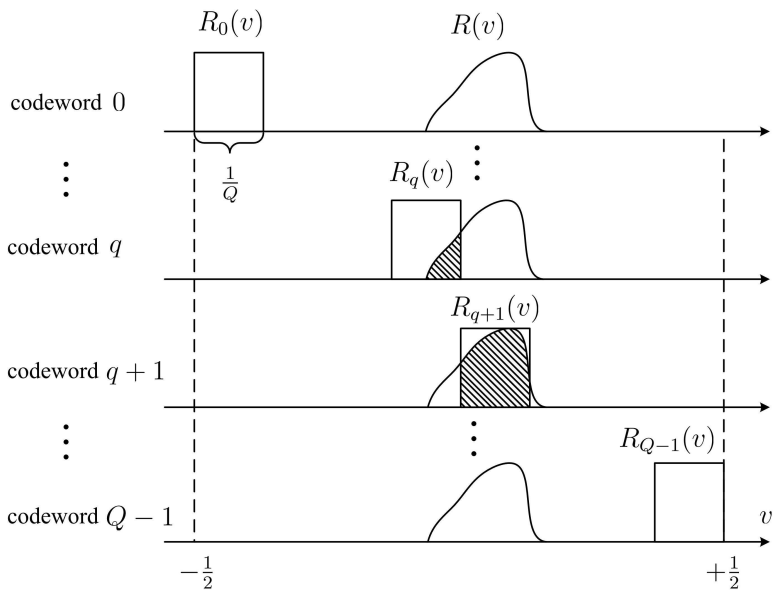}\\
  \caption{Orthogonal codeword spectra used in this paper.}\label{fig:codebook}
\end{figure}

As an alternative, we consider a codebook composed of a group of mutually orthogonal functions. Actually, we can always find such a codebook because the dimension of the functional space is larger than the codebook size. Due to the orthogonality, the chordal distance between any two different codewords will be unit, which has achieved the maximum of (\ref{equ:max_min}). In this sense, a codebook composed of mutually orthogonal functions can be viewed as a good codebook.\par

Although a good codebook can be constructed using a group of orthogonal functions, the best codebook should depend on the knowledge of the AoD distribution, which is usually unknown in advance at the BS. Inspired by the fact that the AoDs are distributed within a single cluster \cite{AAdhikary,AForenza}, as shown in Fig.~\ref{fig:codebook}, we consider the following codebook in this paper,
\begin{align}\label{equ:cb}
R_q(v)=
\begin{cases}
\sqrt{Q}, & -\frac{1}{2}+\frac{q}{Q}\leq v\leq -\frac{1}{2}+\frac{q+1}{Q},\\
0, & \mathrm{otherwise},
\end{cases}.
\end{align}
which divides the whole band into $Q$ equal pieces. The best codeword is then determined by the size of the overlap region between the codeword spectrum and the real spatial spectrum. From \cite{OEdfors}, the dimension of the effective channel can be approximated by $M/Q$.\par
With the codeword spectrum in (\ref{equ:cb}), the corresponding correlation function can be given as
\begin{align}\label{equ:Rq}
r_q[m]&=\int_{-\frac{1}{2}}^{+\frac{1}{2}}R_q(v)e^{j2\pi mv}dv=\frac{1}{\sqrt{Q}}\mathrm{sinc}\left(\frac{\pi m}{Q}\right)e^{j2\pi m(-\frac{1}{2}+\frac{q+0.5}{Q})}.
\end{align}
Based on $r_q[m]$ in (\ref{equ:Rq}), the $q$-th codeword matrix can be obtained. Accordingly, $\mathbf{U}_q$ for the $q$-th codeword can be also generated via EVD. In practical systems, $\mathbf{U}_q$'s can be pre-calculated and stored in the BS. The complexity can be therefore greatly reduced since no need for real-time EVD operation.

\section{Codebook Analysis}
In this section, we first analyze the spatial leakage for the proposed codebook, and then discuss width of the codeword spectrum.

\subsection{Spatial Leakage}
\subsubsection{Infinite $M$}
When $M\rightarrow\infty$, the spatial leakage of the DFT submatrix based inner precoder is reduced to zero \cite{AAdhikary}. Actually, the spatial leakage of the proposed codebook can be also reduced to zero when $M\rightarrow\infty$.\par

Suppose that the $q$-th codeword is fed back to the BS, and thus $\mathbf{U}_q$ is used as the inner precoder. From Appendix B, the columns of the eigen-matrix, $\mathbf{U}_q$, can be viewed as the orthogonal basis functions of a subspace,
\begin{align}
\mathcal{S}=\mathrm{span}\{\mathbf{s}(v)|v\in V_q\},
\end{align}
where $V_q=\{v|-\frac{1}{2}+\frac{q}{Q}\leq v\leq -\frac{1}{2}+\frac{q+1}{Q}\}$ and $\mathbf{s}(v)=\{e^{j2\pi mv}\}_{m=0}^{M-1}\in\mathcal{C}^{M\times 1}$ denotes the steering vector. When the antenna number is very large, we have $\frac{1}{M}\mathbf{s}^{\mathrm{H}}(v_0)\mathbf{s}(v)=0$ for any $v_0\notin V_q$ and $v\in V_q$ \cite{MViberg}. Therefore, $\mathbf{s}(v_0)$ should be orthogonal to any element in the subspace $\mathcal{S}$ since it is spanned by $\{\mathbf{s}(v)|v\in V_q\}$. As a result, $\mathbf{s}(v_0)$ should be also orthogonal to the basis functions of $\mathcal{S}$, that is
\begin{align}\label{equ:signal_space_orth}
\frac{1}{M}\mathbf{s}^{\mathrm{H}}(v)\mathbf{U}_q=\mathbf{0},~\mathrm{for}~\mathbf{s}(v)\notin\mathcal{S},
\end{align}
where $\mathbf{0}$ denotes a $1\times D$ row vector. Denote $A(v)$ as the spectrum of the transmit signal, which is defined as the Fourier transform of the transmit signal over different antennas, that is
\begin{align}\label{Av}
A(v)=\frac{1}{M}\mathbf{s}^{\mathrm{H}}(v)\mathbf{U}_q\mathbf{v}[k]x[k].
\end{align}
Then, using the relation in (\ref{equ:signal_space_orth}), we can obtain $A(v)=0~\mathrm{for}~v\notin V_q$, which means the transmit signal to the undesired direction is zero.\par

\subsubsection{Finite $M$}

In practical systems, the antenna number is always finite, and therefore the spatial leakage cannot be completely mitigated. We will show in this situation that the eigen-matrix based approach causes the smallest leakage among all kinds of inner precoders.\par

For a general inner precoder $\mathbf{W}$, the spectrum of the transmitted signal, similar to (\ref{Av}), can be given by $A(v)=\frac{1}{M}\mathbf{s}^{\mathrm{H}}(v)\mathbf{W}\mathbf{v}[k]x[k]$. For the $q$-th codeword, the spatial leakage of the transmit signal can be minimized if we can find an inner precoder, $\mathbf{W}_{\mathrm{o}}$, such that the signal power of $|A(v)|^2$ most concentrated in $-\frac{1}{2}+\frac{q}{Q}\leq v\leq -\frac{1}{2}+\frac{q+1}{Q}$. Mathematically, this is equivalent to find $\mathbf{W}_{\mathrm{o}}$ to maximize the power concentration
\begin{align}
\lambda(\mathbf{W})=\frac{\int_{-\frac{1}{2}+\frac{q}{Q}}^{-\frac{1}{2}+\frac{q+1}{Q}}|A(v)|^2dv}{\int_{-\frac{1}{2}}^{+\frac{1}{2}}|A(v)|^2dv}.
\end{align}
The solution of the above constrained maximization are the DPS sequences \cite{DSlepian}. The standard DPS sequences are obtained with respect to a symmetrical spectrum, and therefore the standard DPS sequences, $u_{\mathrm{DPS},d}[m]$'s, are defined as the eigenvector of a Hermitian matrix $\left\{\mathrm{sinc}\left[\frac{\pi (m-n)}{Q}\right]\right\}_{m,n=0}^{M-1}$ \cite{TZemen}. In our problem, the desired signal spectrum is modulated by the central wave number of the $q$-th codeword. Accordingly, the DPS sequences should be therefore defined with respect to
\begin{align}
\left\{\mathrm{sinc}\left[\frac{\pi (m-n)}{Q}\right]e^{j2\pi (m-n)(-\frac{1}{2}+\frac{q+0.5}{Q})}\right\}_{m,n=0}^{M-1},
\end{align}
which is exactly the same with the codeword matrix in (\ref{equ:Rq}) except for a scaling constant. As a result, the eigen-vector of the $q$-th codeword matrix can be represented by
\begin{align}
u_{q,d}[m]=u_{\mathrm{DPS},d}[m]e^{-j2\pi m\left(-\frac{1}{2}+\frac{q+0.5}{Q}\right)}.
\end{align}
In this sense, the outer precoder for the proposed codebook can be viewed as modulated DPS sequences. Since the DPS sequences have the best power concentration, the proposed codebook can therefore achieve the smallest spatial leakage among all kinds of inner precoders, including the DFT submatrix based one.

\begin{figure}
  \centering
  \includegraphics[width=4.5in]{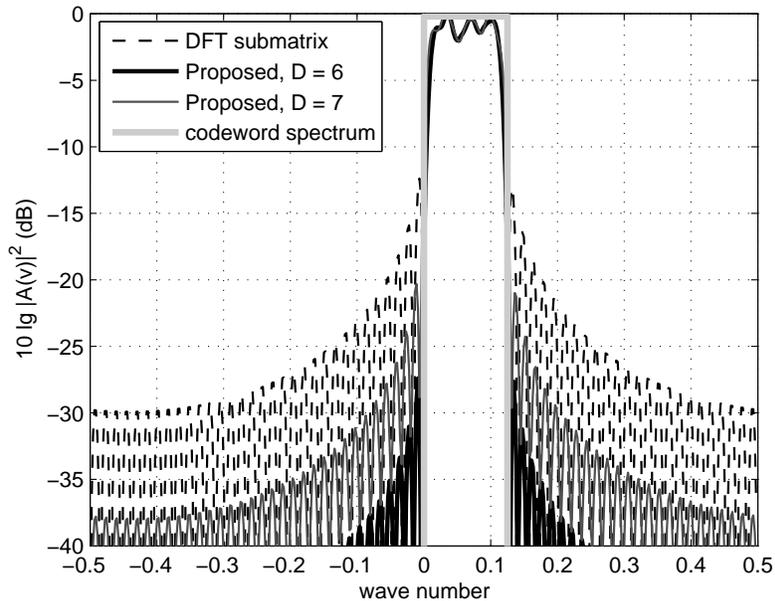}\\
  \caption{A comparison of the transmit spectra for the proposed approach and the DFT submatrix based approach.}\label{fig:spectrum}
\end{figure}

\subsubsection{Example}
As an example in Fig.~\ref{fig:spectrum}, we consider the spectra for an $M=64$ antenna case with $Q=8$. The wave numbers of different paths are assumed to be uniformly distributed within $[0,1/Q]$, corresponding to an AoD range $[0,\arcsin(2/Q)]$. In this case, the codeword spectrum with $q=4$ is the desired inner precoder, and the ideal outer precoders with $D=6\mathrm{~and~}7$ are assumed for both the proposed approach and the DFT submatrix based one. The desired codeword spectrum is also shown in the figure for comparison. As expected, the proposed approach has much lower spatial leakage than the DFT submatrix based precoder. The DPS sequence $u_{q,0}[m]$ has the maximum power concentration, $u_{q,1}[m]$ is the next sequence having the maximum concentration, and so on. Accordingly, the power concentration of $u_{q,5}[m]$ is better than that of $u_{q,6}[m]$, and therefore the case with $D=6$ can achieve lower spatial leakage than the case with $D=7$. The reduction of spatial leakage can not only increase the transmit power on the desired direction but also degrades interferences to the undesired directions. As a result, the performance using the proposed codebook can be improved compared to the DFT submatrix based one, as confirmed by the simulation results.

\subsection{Width of Codeword Spectrum}
The performance of the codebook depends on the width of the codeword spectrum. To gain some insightful analysis, we assume that $D=M/Q$, that is, all dimensions of the effective channel are captured by the outer precoder. Moreover, we also assume that the wave numbers of the paths are integer times of $1/M$ in the subsequential of this section. In this case, $v_l=\frac{n_l}{M}$ where $n_l$'s are integers from $0$ to $M - 1$. As a result, the spatial correlation matrix in (\ref{equ:corr_matrix}) becomes a circular matrix and the eigen-matrix in (\ref{equ:corr_matrix}) can be replaced by a DFT matrix, $\mathbf{F}=\{\frac{1}{\sqrt{M}}e^{-j\frac{2\pi mn}{M}}\}_{m,n=0}^{M-1}$. Correspondingly, $\mathbf{U}_q$ for the $q$-th codeword can be replaced by $\mathbf{F}_q\in\mathcal{C}^{M\times D}$ which is a DFT submatrix including the $\frac{qM}{Q}$-th to the $\left[\frac{(q+1)M}{Q}-1\right]$-th columns of $\mathbf{F}$.

\subsubsection{Average SNR}
Suppose that the $q$-th codeword has the largest average SNR, given by
\begin{align}\label{equ:wide_beam_snr}
\gamma_q&=\mathrm{Tr}(\mathbf{F}_q^{\mathrm{H}}\mathbf{F}\mathbf{\Lambda}\mathbf{F}^{\mathrm{H}}\mathbf{F}_q^{\mathrm{H}})=M\sum_{n_l\in N_q}\sigma_{n_l}^2,
\end{align}
where $N_q=\{m|\frac{qM}{Q}\leq m\leq \frac{(q+1)M}{Q}-1\}$ denotes the range of the wave number covered by the $q$-th codeword spectrum. Apparently, the reduction of $Q$ will increase the width of the codeword spectrum such that more beams can be included for the desired codeword. As a result, the average SNR can be improved by increasing the width of the codeword spectrum because more channel power can be captured.\par

From (\ref{equ:sys_model}), the transmit power is constant for any $\mathbf{W}$ and $\mathbf{v}[k]$ with $\mathbf{W}^{\mathrm{H}}\mathbf{W}=\mathbf{I}$ and $\|\mathbf{v}[k]\|_2^2=1$, that is, $\mathrm{E}(x^*[k]\mathbf{v}^{\mathrm{H}}[k]\mathbf{W}^{\mathrm{H}}\mathbf{W}\mathbf{v}[k]x[k])=1$. Therefore, the width of the codeword spectrum will not affect the transmit power. However, having a smaller $Q$ will increase the dimension of the effective channel. As a result, more array gain can be obtained and thus the average SNR is improved. In other words, the increase of the average SNR by having a smaller $Q$ results from the increased array gain, which can be translated into including more beams.

\subsubsection{Relative SNR Loss}
The relative SNR loss denotes the degradation of the average SNR compared to that of the ideal inner precoder where the spatial covariance matrix is perfectly known at the BS. We will show that increasing the width of the codeword spectrum will almost always cause more relative SNR loss even though it can improve the absolute average SNR.\par

Denote $\gamma_q[D]$ and $\gamma_o[D]$ to be the average SNRs for the proposed approach and the ideal inner precoder both with dimension $D$. Then the relative SNR loss is defined as
\begin{align}
L[D]=\gamma_{o}[D]-\gamma_q[D].
\end{align}
Then, denote $\Delta_q[D]$ and $\Delta_o[D]$ to be the powers of the newly included beams for the proposed approach and the ideal outer precoder,
\begin{align}
\gamma_q[D+1]-\gamma_q[D]=M\Delta_q[D],\label{equ:incre1}\\
\gamma_o[D+1]-\gamma_o[D]=M\Delta_o[D].\label{equ:incre2}
\end{align}
For a set of sorted beam strengths, $\sigma_0^2>\sigma_1^2>\cdots>\sigma_{M-1}^2$, $\Delta_o[D]$ is exactly the $(D+1)$-th entry of this set, that is, $\Delta_o[D]=\sigma_D^2$. For the proposed approach, however, the newly included beam is unknown, and thus $\Delta_q[D]$ is random.\par
To gain insightful results, $\Delta_q[D]$ is assumed to be uniformly distributed within the ordered set above.\footnote{Strictly speaking, this assumption is inaccurate because $D$ paths have been picked out to form the inner precoder. Even though, this assumption can simplify the analysis and we thus adopt it here.} In this case, we have
\begin{align}
&\mathrm{Pr}(\Delta_o[D]\geq\Delta_q[D])=\frac{M-D}{M},\\
&\mathrm{Pr}(\Delta_o[D]<\Delta_q[D])=\frac{D}{M}.
\end{align}
Since $D\ll M$, we therefore almost always have $\Delta_o[D]>\Delta_q[D]$. It means that the newly included beam for the ideal inner precoder is almost always stronger than that of the proposed approach.\par
Subtracting (\ref{equ:incre1}) from (\ref{equ:incre2}),
\begin{align}
L[D+1]-L[D]=M\left(\Delta_o[D]-\Delta_q[D]\right).
\end{align}
As a result, we have $L[D+1]>L[D]$ with high probability. In other words, increasing the dimension of the outer precoder will almost always cause more relative SNR loss. Therefore, a narrow codeword spectrum is required to reduce the relative SNR loss. The above analysis has been also confirmed by the simulation results.\par

In summary, the above analysis shows that widening the codeword spectrum can improve the absolute average SNR but also causes more relative SNR loss compared to the ideal inner precoder.

\section{Practical Consideration}
In this section, we will discuss practical issues of the proposed codebook, including the multiuser case, a downlink training design for SNR estimation, a multi-codeword feedback approach, and the extension to the UPA.
\subsection{Multiuser}
In practical systems, there are usually multiple users in the cell of interest. Even though our codebook is designed with respect to the single-user case, it can be also used when there are multiple users. However, the transmitter at the BS may change, depending on if the users select the same codeword or not.\par

\begin{figure}
  \centering
  \includegraphics[width=2.7in]{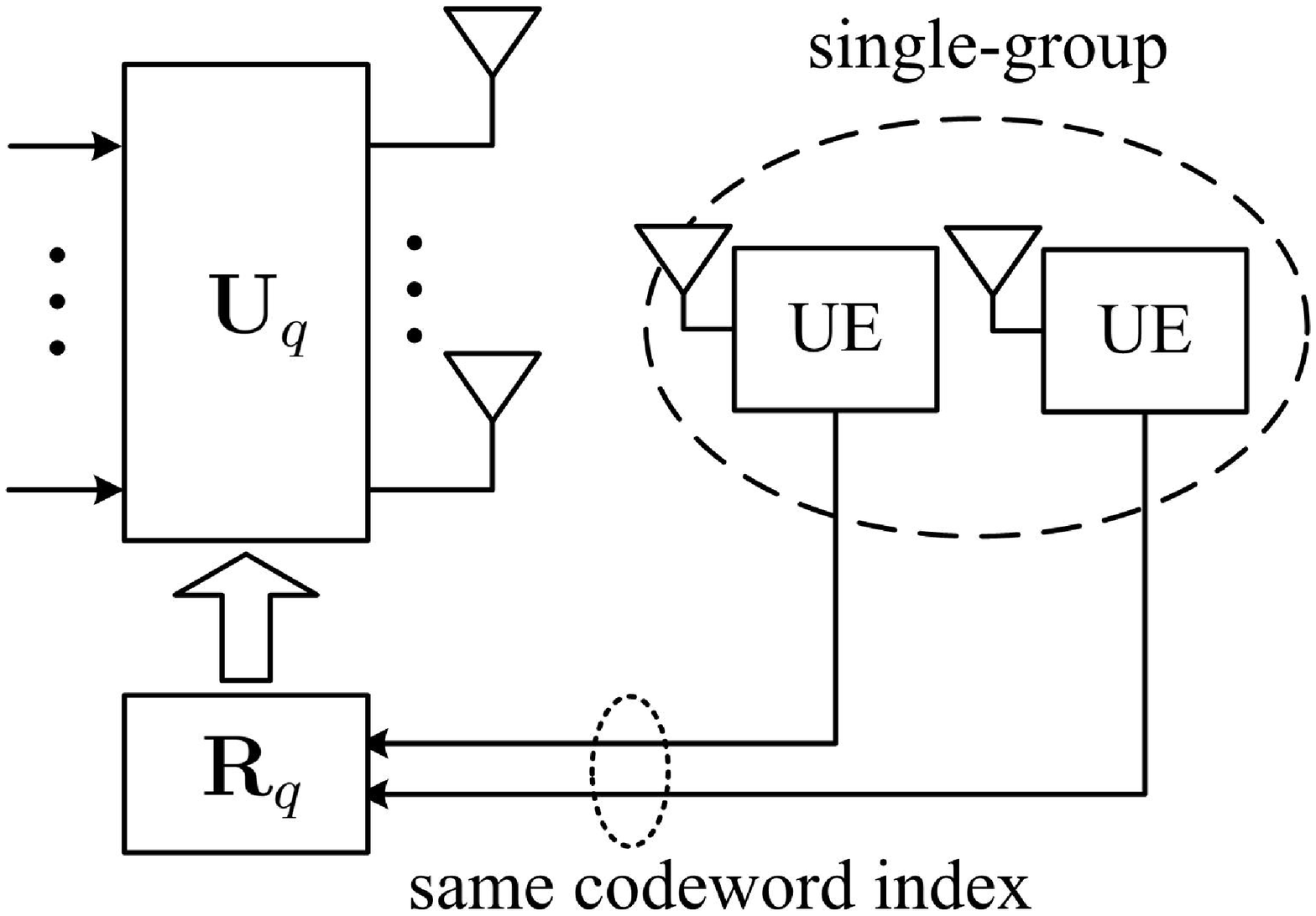}~~
  \includegraphics[width=2.7in]{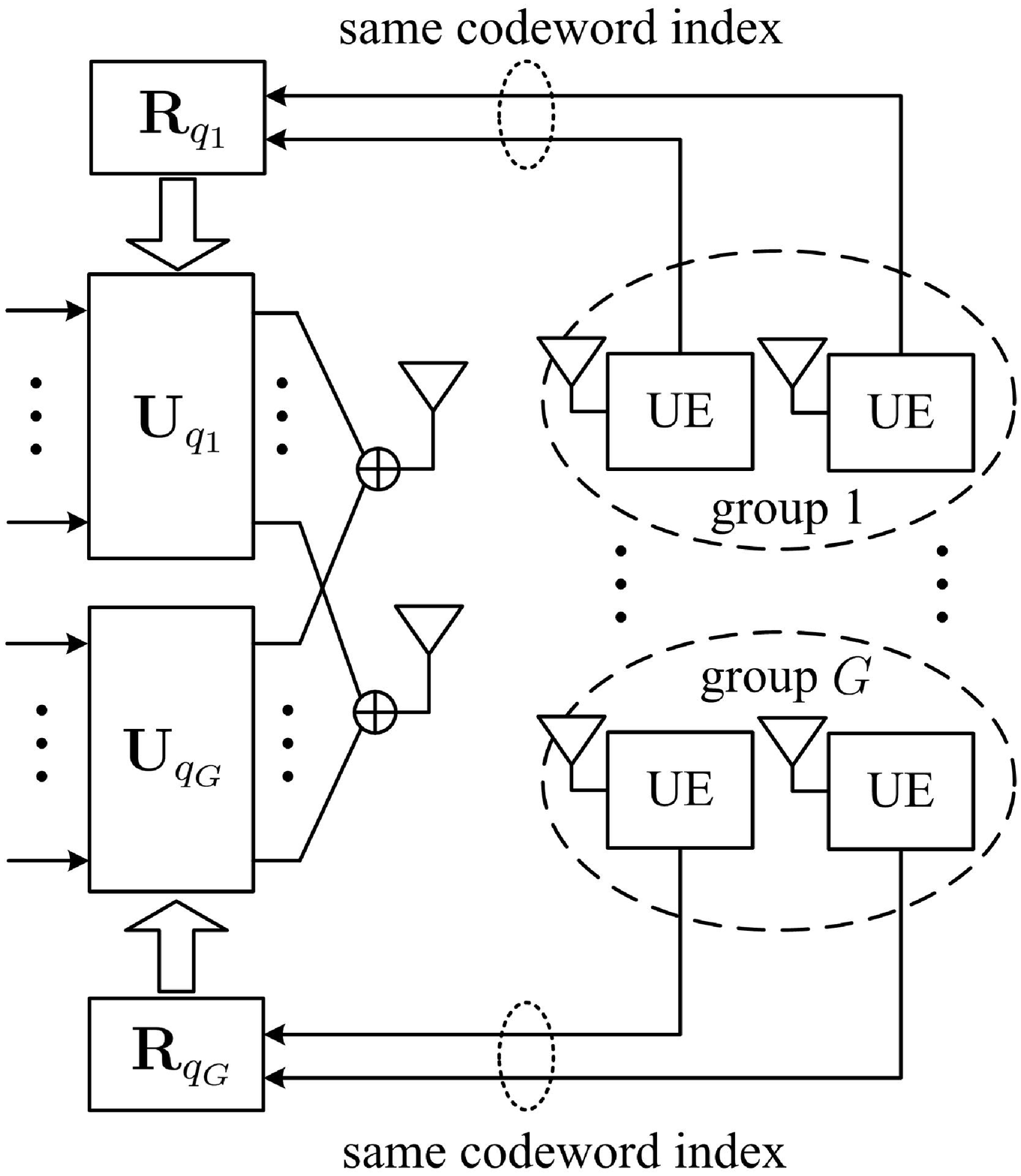}\\
  (a)~~~~~~~~~~~~~~~~~~~~~~~~~~~~~~~~~~~~~~~~(b)
  \caption{(a) single-group multiuser and (b) multi-group multiuser with group $1$ to $G$.}\label{MU}
\end{figure}

\subsubsection{Single-group multiuser}
If the users select the same codeword, then they will share the same inner precoder, as in Fig.~\ref{MU} (a). This is usually the case when the users are close to each other. In this case, the users are within similar scattering environments \cite{IViering}, and thus they will have similar spatial covariance matrices that can be quantized to the same codeword matrix. Since the users share the same inner precoder, they can be viewed as in the same group \cite{AAdhikary}. Accordingly, a zero-forcing (ZF) outer precoder can be used to remove the inter-user-interference (IUI) within the same group.\par

\subsubsection{Multi-group multiuser}
On the other hand, if the users are far from each other, their spatial covariance matrices can be quit different. Correspondingly, those spatial covariance matrices will be quantized to different codewords, and therefore multiple inner precoders should be adopted with each one corresponding to one user group, as in Fig.~\ref{MU} (b) \cite{AAdhikary}. In this case, the system suffers from not only the IUI but also the inter-group-interference (IGI). Although the IGI can be removed using a joint-group processing scheme, it will increase the overhead of CSI estimation and feedback because the BS needs not only the intra-group but also the inter-group CSIs. To reduce the overhead, a simple per-group processing can be adopted \cite{JNam}, which requires only the intra-group CSI for IUI cancellation within a group.  In this situation, the low spatial leakage of the proposed codebook causes much smaller IGI to the undesired groups than the DFT submatrix based one. The performance can be therefore improved, as confirmed by the simulation results.

\subsection{SNR Estimation}
To determine the best codeword, the UE needs to estimate the average SNR for each codeword. In above, we have assumed an ideal outer precoder for the codebook design, which implies that the effective channel is known in advance. In practical systems, the effective channel cannot be determined until the BS determines the inner precoder. As a result, the ideal outer precoder cannot be used for practical downlink training. \par

\begin{figure}
  \centering
  \includegraphics[width=4.5in]{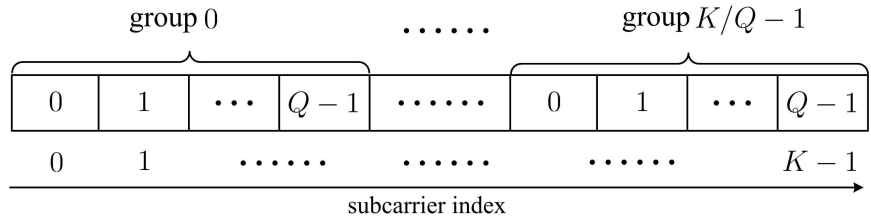}\\
  \caption{Allocation of subcarriers for different codeword matrices.}\label{fig:training}
\end{figure}

To address the above issue, we propose a downlink training approach, from which, the UE can obtain SNR estimations for all codewords with only one training symbol. The downlink training approach consists of a subcarrier allocation strategy for all the codewords and an outer precoder with random coefficients.\par

The subcarrier allocation strategy allows the UE to obtain the average SNR by averaging the instantaneous SNRs over multiple subcarriers. For this purpose, the subcarriers allocated to a codeword should be as far as possible such that the channels at different subcarriers can be viewed as independently distributed. As in Fig.~\ref{fig:training}, we divide all $K$ subcarriers into $K/Q$ groups. Each group has $Q$ subcarriers with each subcarrier allocated to one individual codeword. From the figure, the $q$-th codeword will occupy the subcarriers with indexes $pQ+q$'s where $p=0,1,\cdots,K/Q-1$. In other words, two adjacent subcarriers allocated to the same codeword are separated by $Q$ subcarriers.

For $D=1$, the SNR can be estimated using any scalar outer precoder with unit power. It has been shown in Section III that the SNR depends on the effective channel for $D>1$, which cannot be known until the BS determines the inner precoder. In the subsequence of this section, we will show that a random precoder can also estimate the SNR by averaging the received signals over different subcarriers. For this purpose, the random precoder can be generated such that
\begin{align}\label{equ:precoder}
\mathrm{E}(\mathbf{v}[pQ+q]\mathbf{v}^{\mathrm{H}}[p_1Q+q])=\delta[p-p_1]\mathbf{I},
\end{align}
where the gain has been normalized for simple notation. There is no particular requirement on the distribution of those random coefficients and we therefore assume they are generated from a quadrature-phase-shift-keyying (QPSK) constellation for simplicity.\par

With the subcarrier allocation and the random outer precoder, the estimation of the average SNR for the $q$-th quantized covariance matrix can be given by
\begin{align}\label{equ:snr_est}
\widehat{\gamma}_q&=\frac{Q}{K}\sum_{p=0}^{\frac{K}{Q}-1}\left|x^*[pQ+q]y[pQ+q]\right|^2\nonumber\\
&=\frac{Q}{K}\sum_{p=0}^{\frac{K}{Q}-1}\left\{|\mathbf{h}^{\mathrm{H}}[pQ+q]\mathbf{U}_q\mathbf{v}[pQ+q]|^2+|z[pQ+q]|^2+\right.\nonumber\\
&~~~~~~~~~~~~\left.2\mathrm{Re}(\mathbf{h}^{\mathrm{H}}[pQ+q]\mathbf{U}_q\mathbf{v}[pQ+q]x^*[pQ+q]z[pQ+q])\right\},
\end{align}
where $x[pQ+q]$'s denotes the training symbols. If the number of subcarriers allocated to each codeword is large enough, the sample average in (\ref{equ:snr_est}) can be approximated by the statistical average. From Appendix C, we have
\begin{align}
&\mathrm{E}(|\mathbf{h}^{\mathrm{H}}[pQ+q]\mathbf{U}_q\mathbf{v}[pQ+q]|^2)=\mathrm{Tr}(\mathbf{U}_q^{\mathrm{H}}\mathbf{R}\mathbf{U}_q),\label{equ:term1}\\
&\mathrm{E}(|z[pQ+q]|^2)=N_0,\label{equ:term2}\\
&\mathrm{E}(\mathbf{h}^{\mathrm{H}}[pQ+q]\mathbf{U}_q\mathbf{v}[pQ+q]x^*[pQ+q]z[pQ+q])=0,\label{equ:term3}
\end{align}
where we have used the identity (\ref{equ:precoder}) for (\ref{equ:term1}). Substituting (\ref{equ:term1}) to (\ref{equ:term3}) into (\ref{equ:snr_est}), the SNR estimation is then reduced to
\begin{align}\label{equ:snr_accuracy}
\widehat{\gamma}_q=\gamma_q+N_0.
\end{align}
Since the noise power, $N_0$, is constant for different codewords, selection of the codeword using the SNR estimation is then equivalent to that using the true SNR.\par

The mean-square-error (MSE) for the SNR estimation can be defined as
\begin{align}\label{equ:MSE}
\mathrm{MSE}=\frac{1}{Q}\sum_{q=0}^{Q-1}\mathrm{E}\left(\frac{|\widehat{\gamma}_q-(\gamma_q+N_0)|^2}{|\gamma_q+N_0|^2}\right).
\end{align}
It is in general difficult to derive a closed form for (\ref{equ:MSE}) because the higher order statistics make the calculation intractable. We therefore evaluate the MSE in the simulation as in Section VI.
\subsection{Multi-Codeword Feedback}
In Section III, our codebook is designed based on the assumption that the AoDs are within a single cluster \cite{AAdhikary,AForenza}, which may not be the case for some practical channels. On the other hand, the analysis in Section IV shows that widening the codeword spectrum can improve the SNR but also causes more relative SNR loss compared to the ideal precoder.\par

To address the above issue and achieve a tradeoff, we can adopt a relatively narrower codeword spectrum by using a larger $Q$, and then feed back multiple codewords to the BS. The narrower codeword spectrum guarantees that the relative SNR loss can be kept small, while feedback of multiple codewords can maintain the captured channel power unchanged. In fact, when the AoDs are distributed within multiple clusters, feedback of multiple codewords can obtain the channel power more efficiently so that the SNR can be even improved.


\subsection{UPA}
In above, we only consider the simple ULA to simplify the derivation and the analysis. In practical systems, UPA is more widely used because more antennas can be installed within a smaller area. Even though our approach is proposed based on the ULA, it can be also extended to the UPA case in a straightforward way.\par

For UPA antenna, the array is composed of $M_{\mathrm{V}}\times M_{\mathrm{H}}$ antennas, where $M_{\mathrm{V}}$ and $M_{\mathrm{H}}$ denote the numbers of antennas in the vertical direction and horizontal direction, respectively. Denote $h_{m,n}[k]$ with $0\leq m\leq M_{\mathrm{V}}-1$ and $0\leq n\leq M_{\mathrm{H}}-1$ to be the channel at the $(m,n)$-th antenna corresponding to the $k$-th subcarrier, then the two-dimensional correlation function, similar to (\ref{equ:sp_corr}), can be defined as $r[m,n]\triangleq\mathrm{E}(h_{m_0+m,n_0+n}[k]h_{m_0,n_0}^*[k])$. Similar to (\ref{equ:cb}), the codebook can be obtained by dividing the two-dimensional wave number domain into $P\times Q$ zones with each zone corresponding to a two-dimensional codeword spectrum. Correspondingly, the $(p,q)$-th quantized matrix ($0\leq p\leq P-1,0\leq q\leq Q-1$) can be given by
\begin{align}
r_{p,q}[m,n]=\frac{1}{\sqrt{PQ}}\mathrm{sinc}\left(\frac{\pi m}{P}\right)\mathrm{sinc}\left(\frac{\pi n}{Q}\right)e^{-j2\pi m(-\frac{1}{2}+\frac{p+0.5}{P})}e^{-j2\pi n(-\frac{1}{2}+\frac{q+0.5}{Q})},
\end{align}
which can be viewed as a modulated two-dimensional DPS sequence. As a result, low spatial leakage can be still achieved compared to the two-dimensional DFT submatrix based approach.

\section{Simulation Results}
In this section, computer simulation is conducted to verify the proposed approach. In the simulation, we consider an OFDM modulation with $15$ KHz subcarrier spacing. A typical ULA with $M=64$ antennas spaced by the half wave-length is considered and the size of the outer precoder is $D=6$. To evaluate the proposed approach in practical systems, we use the parameters in WINNER II channel model (C2 scenario for typical urban macro-cell without line-of-sight (LOS) path) for simulation \cite{WINNER}, where the channel is composed of $20$ multipath clusters with each cluster having $20$ subpaths. The path delays and the corresponding power delay profile are generated following the distribution defined in WINNER II channel model. The AoDs depend on the LOS direction of the user, which is determined by the user coordinates with respect to the position of the BS. In the simulation, we assume the LOS direction angles of the different users are uniformly distributed in $(-85^{\mathrm{o}},+85^{\mathrm{o}})$.

\begin{figure}
  \centering
  \includegraphics[width=4.5in]{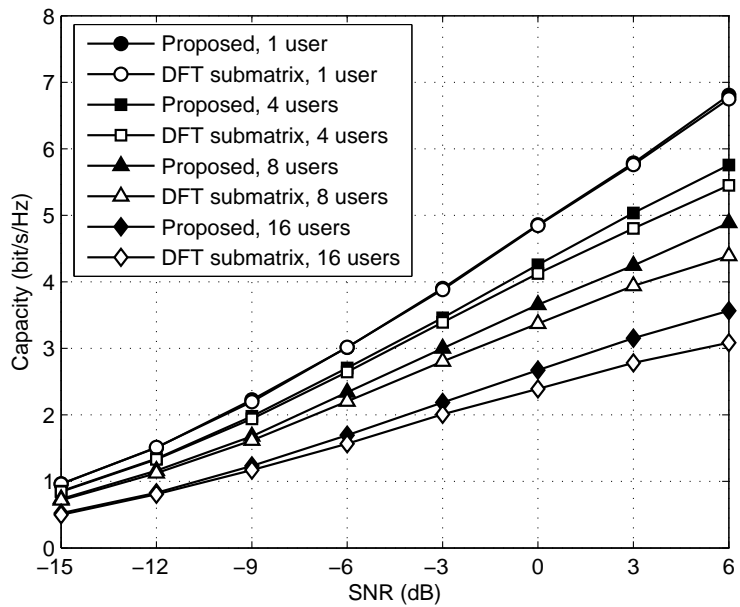}\\
  \caption{Capacities versus SNR for different user numbers.}\label{fig_MU}
\end{figure}

Fig.~\ref{fig_MU} shows the average capacity over different users versus SNR with $Q=8$. As a comparison, the performance of the DFT submatrix based inner precoder is also shown in the figure. From Fig.~\ref{fig_MU}, the proposed codebook has almost the same performance with the DFT submatrix based one because most of the power can be captured within the main lobes of both approaches even though the proposed codebook can achieve much lower spatial leakage. On the other hand, significant improvement can be observed for the proposed codebook when there are multiple users, especially when the user number is large. Since the spatial leakage of the proposed codebook is much lower, the IGI can be greatly reduced compared to the DFT submatrix based approach. This is especially the case for the large user numbers because IGI is more severe in this situation and thus the reduction of IGI will result in more significant improvement.

\begin{figure}
  \centering
  \includegraphics[width=4.5in]{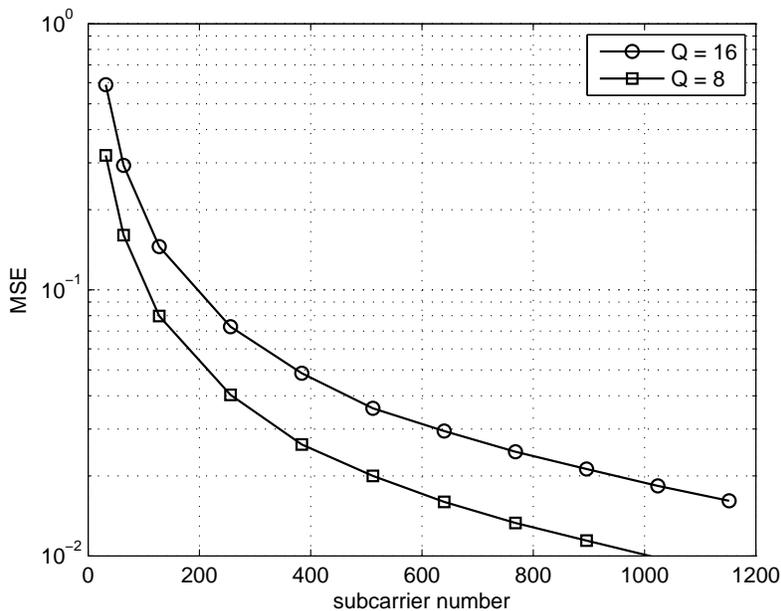}\\
  \caption{MSE of the SNR estimation for different numbers of quantized covariance matrices.}\label{fig_MSE}
\end{figure}
Fig.~\ref{fig_MSE} shows the MSE of the SNR estimation for different numbers of quantized matrices, where we have assumed that the channels at the subcarriers allocated to each codeword are independent. From the figure, a smaller $Q$, corresponding to a wider codeword spectrum, can achieve better estimation performance. For a smaller $Q$, more subcarriers can be allocated to a codeword such that the average can be conducted with more samples, leading to the improved estimation accuracy.\par

\begin{figure}
  \centering
  \includegraphics[width=4.5in]{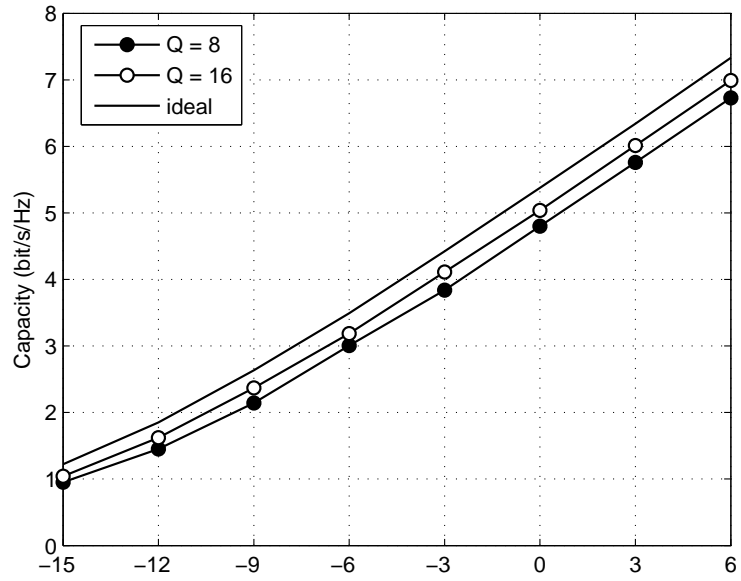}\\
  \caption{Comparison between single-codeword feedback and multi-codeword feedback.}\label{fig_multicw}
\end{figure}

Fig.~\ref{fig_multicw} shows the relative SNR loss of different sizes of the codeword spectra. When $Q=16$, two codewords are fed back with each one corresponding to an outer precoder of dimension $D=3$, such that the total dimension of the outer precoder is still six. Only the single user is considered in this figure. From the figure, a smaller $Q$ suffers from more relative SNR loss, which coincides with our analysis in Section VI. With a large probability, the newly included beam for the ideal precoder is stronger than that for the proposed approach, and thus the relative SNR loss is increased for a wider codeword spectrum.


\begin{figure}
  \centering
  \includegraphics[width=4.5in]{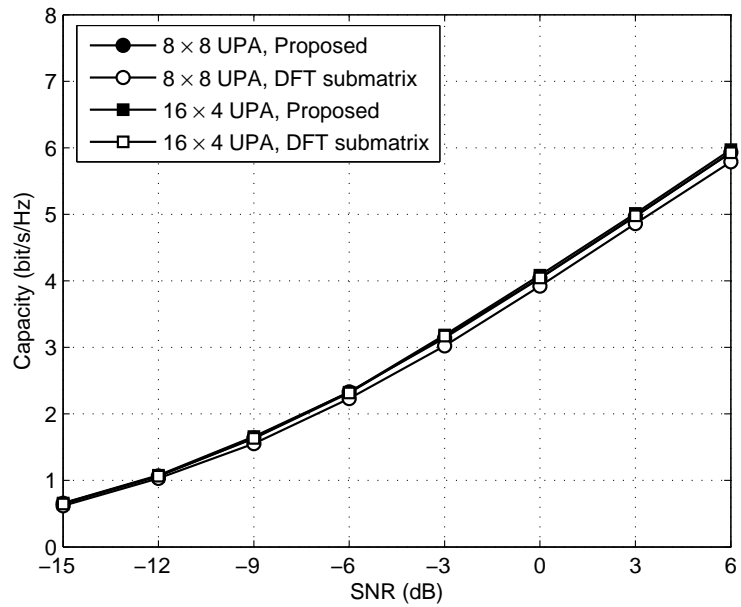}\\
  \caption{Performances of the UPA with different antenna configurations.}\label{fig_UPA}
\end{figure}

For the UPA case, the spatial covariance matrix is not only determined by the azimuth AoDs but also the zenith AoDs. To take various situation into account, the azimuth and zenith AoDs are assumed independently and uniformly distributed within a range that has a random central angle and a random angular spread uniformly distributed within $(-180^{\mathrm{o}},+180^{\mathrm{o}})$ and $(0^{\mathrm{o}},90^{\mathrm{o}})$, respectively. For the simulation results in Fig.~\ref{fig_UPA}, we consider the single-user case, $(P,Q)=(4,2)$, and two antenna configurations with $(M,N)=(8,8)$ and $(M,N)=(16,4)$. As a comparison, the two-dimensional DFT submatrix based approach is also considered in the simulation. From the figure, the proposed codebook can achieve a small performance improvement over the DFT submatrix based approach, which coincides with the results for the ULA case.

\section{Conclusions}
In this paper, we have investigated the the quantization and the feedback for the spatial covariance matrix in massive MIMO systems with cascaded precoding. Our analysis has shown that the proposed approach can achieve much smaller spatial leakage than the traditional DFT submatrix based approach. Practical issues, including multiuser scenario, SNR estimation, the multi-codeword feedback scheme, and UPA case, have also been addressed in this paper. Our simulation results have shown that the proposed approach can achieve significant performance improvement over the DFT submatrix based approach, especially when the user number is large.

\appendices
\renewcommand{\theequation}{A.\arabic{equation}}
\setcounter{equation}{0}
\section{}
For any non-zero vector $\mathbf{w}\in\mathcal{C}^{D\times 1}$, we have
\begin{align}
\mathrm{E}(|\mathbf{w}^{\mathrm{H}}\mathbf{U}_q^{\mathrm{H}}\mathbf{h}[k]|^2)&=\mathrm{E}(\mathbf{w}^{\mathrm{H}}\mathbf{U}_q^{\mathrm{H}}\mathbf{h}[k]\mathbf{h}^{\mathrm{H}}[k]\mathbf{U}_q\mathbf{w})=\mathbf{w}^{\mathrm{H}}\mathbf{U}_q^{\mathrm{H}}\mathbf{R}\mathbf{U}_q\mathbf{w}.
\end{align}
Since $\mathrm{E}(|\mathbf{w}^{\mathrm{H}}\mathbf{U}_q\mathbf{h}[k]|^2)\geq 0$, $\mathbf{w}^{\mathrm{H}}\mathbf{U}_q\mathbf{R}\mathbf{U}_q\mathbf{w}\geq 0$ holds for any non-zero vector, $\mathbf{w}$. As a result, $\mathbf{U}_q\mathbf{R}\mathbf{U}_q$ is a semi-positive definite matrix, and thus $[\mathbf{U}_q^{\mathrm{H}}\mathbf{R}\mathbf{U}_q]_{(m,m)}\geq 0$ for $0\leq m\leq D-1$. Therefore,
\begin{align}
\mathrm{Tr}(\mathbf{U}_q^{\mathrm{H}}\mathbf{R}\mathbf{U}_q)&=\sum_{m=0}^{D-1}[\mathbf{U}_q^{\mathrm{H}}\mathbf{R}\mathbf{U}_q]_{(m,m)}\nonumber\\
&=\lambda_{q,\mathrm{min}}^{-1}\sum_{m=0}^{D-1}\lambda_{q,\mathrm{min}}[\mathbf{U}_q^{\mathrm{H}}\mathbf{R}\mathbf{U}_q]_{(m,m)}\nonumber\\
&\leq \lambda_{q,\mathrm{min}}^{-1}\sum_{m=0}^{D-1}\lambda_{q}^{\frac{1}{2}}[m][\mathbf{U}_q^{\mathrm{H}}\mathbf{R}\mathbf{U}_q]_{(m,m)}\lambda_{q}^{\frac{1}{2}}[m]\nonumber\\
&=\lambda_{q,\mathrm{min}}^{-1}\mathrm{Tr}(\mathbf{\Lambda}_q^{\frac{1}{2}}\mathbf{U}_q^{\mathrm{H}}\mathbf{R}\mathbf{U}_q\mathbf{\Lambda}_q^{\frac{1}{2}}).
\end{align}
where we have used the fact $\lambda_q[m]>0$. Then, using $\mathrm{Tr}(\mathbf{AB})=\mathrm{Tr}(\mathbf{BA})$ where $\mathbf{A}$ and $\mathbf{B}$ are two arbitrary matrices, we obtain
\begin{align}
\mathrm{Tr}(\mathbf{U}_q^{\mathrm{H}}\mathbf{R}\mathbf{U}_q)\leq \lambda_{q,\mathrm{min}}^{-1}\mathrm{Tr}(\mathbf{R}_q\mathbf{R}).
\end{align}
Similarly, we can obtain
\begin{align}
\lambda_{q,\mathrm{max}}^{-1}\mathrm{Tr}(\mathbf{R}_q\mathbf{R})\leq\mathrm{Tr}(\mathbf{U}_q^{\mathrm{H}}\mathbf{R}\mathbf{U}_q).
\end{align}

\renewcommand{\theequation}{B.\arabic{equation}}
\setcounter{equation}{0}
\section{}
Denote $\mathbf{S}_N$ to be an $M\times N$ matrix,
\begin{align}
\mathbf{S}_N=\left\{\mathbf{s}\left(-\frac{1}{2}+\frac{q}{Q}+\frac{n}{QN}\right)\right\}\Big|_{n=0}^{N-1}.
\end{align}
Then the subspace $\mathcal{S}$ can be viewed as the column space of $\mathbf{S}_{\infty}$. Therefore, the orthogonal basis functions of subspace $\mathcal{S}$ is determined by the eigen-matrix of the EVD for $\frac{1}{N}\mathbf{S}_N\mathbf{S}_N^{\mathrm{H}}$ when $N\rightarrow\infty$. On the other hand, when $N\rightarrow\infty$, we have
\begin{align}\label{B3}
\lim_{N\rightarrow\infty}\frac{1}{N}\mathbf{S}_{N}\mathbf{S}_N^{\mathrm{H}}=Q\int_{-\frac{1}{2}+\frac{q}{Q}}^{-\frac{1}{2}+\frac{q+1}{Q}}\mathbf{s}(v)\mathbf{s}^{
\mathrm{H}}(v)dv = Q\mathbf{R}_q,
\end{align}
where the second equation can be easily verified using (\ref{equ:Rq}). Therefore, the orthogonal basis functions of subspace $\mathcal{S}$ is determined by the the eigen-matrix of $\mathbf{R}_q$, which is exactly $\mathbf{U}_q$.
\setcounter{equation}{0}
\renewcommand{\theequation}{C.\arabic{equation}}
\section{}
The left side of (\ref{equ:term1}) can be represented as
\begin{align}\label{C1}
\mathrm{E}(|\mathbf{h}^{\mathrm{H}}[pQ+q]\mathbf{U}_q\mathbf{v}[pQ+q]|^2)&=\sum_{d_1=0}^{D-1}\sum_{d_2=0}^{D-1}\sum_{m_1=0}^{M-1}\sum_{m_2=0}^{M-1}u_{q,d_1}[m_1]u_{q,d_2}^*[m_2]\cdot\nonumber\\
&~~~~\mathrm{E}(h_{m_1}^*[pQ+q]v_{d_1}[pQ+q]h_{m_2}[pQ+q]v_{d_2}^*[pQ+q]),
\end{align}
where $u_{q,d}[m]$ denotes the $m$-th entry on the $d$-th column of matrix $\mathbf{U}_q$ and $v_{d}[pQ+q]$ denotes the $d$-th entry of the outer precoder $\mathbf{v}[pQ+q]$. Note that the statistical average in (\ref{C1}) is with respect to both the channels and the precoding coefficients. Then, using the relation between the cumulant and the expectation \cite{CLNikias}, we have
\begin{align}\label{C2}
&\mathrm{E}(h_{m_1}^*[pQ+q]v_{d_1}[pQ+q]h_{m_2}[pQ+q]v_{d_2}^*[pQ+q])\nonumber\\
=&\mathrm{E}(h_{m_1}^*[pQ+q]h_{m_2}[pQ+q])\mathrm{E}(v_{d_1}[pQ+q]v_{d_2}^*[pQ+q]),
\end{align}
since $v_{d}[pQ+q]$'s and $h_{m}[pQ+q]$'s are mutually independent. Meanwhile, we can obtain from (\ref{equ:sp_corr}) and (\ref{equ:precoder}) that
\begin{align}
&\mathrm{E}(h_{m_1}^*[pQ+q]h_{m_2}[pQ+q])=r[m_2-m_1],\\
&\mathrm{E}(v_{d_1}[pQ+q]v_{d_2}^*[pQ+q])=\delta[d_1-d_2],
\end{align}
we therefore have
\begin{align}\label{C4}
\mathrm{E}(|\mathbf{h}^{\mathrm{H}}[pQ+q]\mathbf{U}_q\mathbf{v}[pQ+q]|^2)&=\sum_{d=0}^{D-1}\sum_{m_1=0}^{M-1}\sum_{m_2=0}^{M-1}u_{q,d}[m_1]u_{q,d}^*[m_2]r[m_2-m_1]\nonumber\\
&=\sum_{d=0}^{D-1}\mathbf{u}_{q,d}^{\mathrm{H}}\mathbf{R}\mathbf{u}_{q,d},
\end{align}
where $\mathbf{u}_{q,d}$ denotes the $d$-th column of matrix $\mathbf{U}_q$. Rewrite the second equation in (\ref{C4}) in a matrix form will lead to the result in (\ref{equ:term1}).\par
Using similar approach, we can easily obtain (\ref{equ:term2}) and (\ref{equ:term3}) since the channels, the precoding coefficients, the transmit symbols, and the additive noise are also mutually independent.
\bibliographystyle{IEEEtran}
\bibliography{IEEEabrv,lsmimobib}

\end{document}